\documentclass[12pt,preprint]{aastex}
\newcommand\xte{XTE J1810$-$197}
\newcommand\xmm{{\it XMM-Newton}}
\newcommand\chandra{{\it Chandra}}

\shorttitle{The Evolution of \xte}
\shortauthors{Alford \& Halpern}
\begin{document}
\title{Evolution of the X-ray Properties of the Transient Magnetar \xte}
\author{
J. A. J. Alford and J. P. Halpern
}
\affil
{\rm Columbia Astrophysics Laboratory, Columbia University, 550 West 120th Street, New York NY, 10027, USA; jason@astro.columbia.edu}

\begin{abstract}
We report on X-ray observations of the 5.54 s transient magnetar \xte \ using the \xmm \ and \chandra \ observatories,
analyzing new data from 2008 through 2014,
and re-analyzing data from 2003 through 2007 with the benefit of these six years of new data.
From the discovery of \xte \ during its 2003 outburst to the most recent 2014 observations, its $0.3-10$ keV X-ray flux has declined by a factor of about 50 from $4.1 \times 10^{-11}$ to $8.1 \times 10^{-13}$ erg cm$^{-2}$ s$^{-1}$.
Its X-ray spectrum has now reached a steady state.
Pulsations continue to be detected from a 0.3 keV thermal hot-spot that remains on the neutron star surface. 
The luminosity of this hot-spot exceeds \xte's spin down luminosity, indicating continuing magnetar activity.
We find that \xte's X-ray spectrum is best described by a multiple component blackbody model in which the coldest 0.14 keV component likely originates from the entire neutron star surface, and the thermal hot-spot is, at different epochs, well described by an either one or two-component blackbody model. 
A 1.2 keV absorption line, possibly due to resonant proton scattering, is detected at all epochs.
The X-ray flux of the hot spot decreased by $\approx 20 \%$ between 2008 March and 2009 March, the same period during which \xte \ became radio quiet.
\end{abstract}

\section{Introduction}

Magnetars are a subclass of neutron stars (NSs) that have long spin periods (2--12 s) and large inferred dipole
magnetic fields ($\approx 10^{14-15} $ G).  They are also characterized by unsteady spin down and variable X-ray emission. 
Unlike ``normal'' rotation powered radio pulsars, magnetars' X-ray luminosity can exceed their spin-down power.
This luminosity is thought to be supplied by their decaying magnetic fields. 

\xte, the first discovered transient magnetar, was found during its X-ray outburst in 2003 \citep{ibr04}.
Subsequently \xte \ became the first magnetar observed to produce emission at radio frequencies \citep{hal05,cam06}. 
Currently, four of the twenty-three confirmed magnetars are known to be transient radio emitters \citep{ola14}.
The X-ray flux of \xte declined over the next four years, and it become radio quiet in late 2008 \citep{got07,ber09,cam06,cam15}.  

A theoretical model of magnetar outbursts relevant to \xte \ has been developed by \cite{bel09}.  
In Beloborodov's model, a twisted NS magnetosphere stores energy that is released in the X-ray outburst.  
As the magnetosphere untwists, particles impact and heat the NS surface, resulting in observable thermal emission.
Beloborodov's model predicts an approximately exponential luminosity decay consistent with early observations of \xte.
Models of magnetar spectra relevant to \xte \ have been developed by \cite{guv07}, \cite{alb10}, and 
\cite{ber11}.
Guver et al. calculated the effects of photon propagation through the NS atmosphere and magnetosphere, accounting for both quantum and general relativistic effects.
Guver et al. fit their model to \xte \ data and calculated a magnetospheric magnetic field strength close to the dipole spin-down value.
Bernardini et al. calculated the effects of light bending and relativistic beaming on the X-ray spectrum.  
They fit this model to multiple epochs of data and calculated the viewing geometry of hotter regions on the surface of \xte.
Albano et al. used Monte Carlo simulations of the propagation of photons from the NS surface through the magnetosphere and produced synthetic pulse profiles to fit to observations.
Albano et al. found that a limited region of the surface of \xte \ was heated, and that this region's size and temperature were decreasing after its outburst.

In this paper we report on the long term X-ray evolution of \xte, with all available \xmm \ data including previously unpublished data from 2008 through 2014.
We also analyze a \chandra \ observation taken near the time when \xte \ became radio quiet in late 2008, looking for correlations between its X-ray and radio activity.
We explore several spectral models, and attempt to determine if \xte \ has returned to a ``quiescent'' state.

\section{Data Reduction and Analysis}

Table 1 summarizes the X-ray observations considered in this paper.  
We mostly focus on the \xmm \ observations because of \xmm's broader energy coverage, larger effective area, and more stable effective area. 

\begin{table}
\caption{Log of \xmm \ Observations}
\begin{tabular}{lccc}
\hline \hline
ObsID  &  Date  & Exposure & Net Counts \\
 & (UT) & (ks) & (0.3 - 10 keV) \\
\hline
0161360301 & 2003 Sep 8 & 12.1 &  68958  \\
0152833201 & 2003 Oct 12  & 8.9 & 29268  \\
0161360501 & 2004 Mar 11  & 18.9 &  15485 \\                   
0164560601 & 2004 Sep 18    &  28.9 & 83341 \\
0301270501 & 2005 Mar 18    &  42.2  & 64978 \\
0301270401 & 2005 Sep 20   &  42.2 & 28726 \\
0301270301 & 2006 Mar 12   &  51.4 & 14104 \\
0406800601 & 2006 Sep 24   &  50.3 & 21107 \\
0406800701 & 2007 Mar 6   &  68.3 & 17347 \\
0504650201 & 2007 Sep 16  & 74.9 & 31397  \\
7594 (\chandra)     &  2008 Mar 18   &  29.6 & 5786 \\
0552800201 & 2009 Mar 5    &  65.8 & 14022  \\
0605990201 & 2009 Sep 5  &  21.6 & 7746 \\
0605990301 & 2009 Sep 7   &  19.9 & 6738 \\
0605990401 & 2009 Sep 23  & 14.2 & 4666 \\
0605990501 & 2010 Apr 9  &  9.9 & 1485  \\
0605990601 & 2010 Sep 5 &   11.3 & 3655 \\
0671060101 & 2011 Apr 3  &   22.9 & 6975  \\
0671060201 & 2011 Sep 9 &   15.9 & 5206 \\
0691070301 & 2012 Sep 6  &  17.9 & 6335 \\
0691070401 & 2013 Mar 3  & 17.9 & 3170 \\
0720780201 & 2013 Sep 5  & 24.5 & 7857 \\
0720780301 & 2014 Mar 4  & 26.0 & 8200 \\
\hline
\end{tabular}
\end{table}

\subsection{Data Reduction}

All of the XMM-Newton data were reduced and extracted using the Science Analysis Software (SAS) version 13.5.  
We only used data from the EPIC pn CCD because of its superior long term stability and throughput at low energies (which was particularly important in the later observations as the blackbody temperatures decreased). 
All observations were performed in Imaging Large Window mode, with the exception of the 2003 September 8 and 2003 October 12 observations, which were performed in Small Window mode and Full Window mode respectively.
The XMM-Newton data were filtered to remove time intervals with high-energy flares, and events with FLAG $ = 0$ and PATTERN $\le4$ were selected from these good time intervals.  
Circular source and background regions were created with radii of $45^{''}$, or sometimes as small as $30^{''}$ in order to avoid overlap with the edge of a detector chip.  
The spectra were grouped with a minimum of 25 counts per bin, and such that the energy resolution was oversampled by no more than a factor of 3.  

The \chandra \ observation of 2008 March 18 was reduced and extracted using the Chandra Interactive Analysis of Observations software (CIAO) version 4.6. 
The observation was performed with the Advanced CCD Imaging Spectrometer spectral component (ACIS-S) in Very Faint Timed Exposure mode, with a custom subarray
of 100 rows to achieve a frame time of 0.3~s, sufficient to resolve the 5.54~s pulsations.
We followed the online CIAO thread\footnote{http://cxc.harvard.edu/ciao/threads/pointlike/} ``Extract Spectrum and Response Files for a Point-like Source''.
Circular source and background regions were created with radii of $8^{''}$.
The source spectrum was grouped with the default value of 15 counts per bin.

All spectral fitting of the \xmm \ and \chandra \ data was performed using XSPEC \citep{arn96}.

\subsection{Three-To-Two-Blackbody Model}

Following \cite{ber09}, we found that the first seven \xmm \ observations, from 2003 September 8 to  2006 March 12, are well described by a three blackbody model, 
where the lowest temperature component (the ``cold'' region) is interpreted as emission from the whole NS surface.  
The temperature and area of this cold component are therefore held constant across all epochs.  
The hot temperature component is thought of as a small hot spot on the NS surface, and the warm temperature component is thought of as warm annulus surrounding the hot spot.  
This simplified model is a rough approximation to hot spot with a large temperature gradient on the surface of the NS.  
This temperature gradient is determined by the details of the heating and the NS surface thermal conductivity. 
The central region is cooling faster than the cooler outer regions, and there is gradual transition to times when the hot spot and annulus are well described by a single temperature.
Starting with the eighth XMM observation on 2006 September 24, and through the most recent observation on 2014 March 4, we find that the X-ray spectra are well modeled by a two temperature blackbody model. 
The coldest temperature component is again interpreted as emission from the whole surface and is therefore held constant across all epochs (including the earlier observations from 2003 September 8 to  2006 March 12).  

This model, in which the spectra of \xte \ evolve from being well described by three blackbodies at earlier times to two blackbodies at later times, is what we refer to here as the three-to-two-blackbody model.
This is distinct from the two blackbody model, described in the next subsection, in which the spectra are fitted by two blackbodies at all epochs, and it is not assumed that we can see emission from the whole stellar surface.
Essential features of the three-to-two-blackbody model are that hydrogen column density, cold blackbody temperature and cold blackbody normalization are held constant at all epochs.
The hot and warm blackbody components are allowed to vary independently at each epoch.
Another feature of the three-to-two-blackbody model is an absorption line at 1.2 keV.

The details of the fitting procedure are as follows.  
We began by simultaneously fitting the last fifteen \xmm \ observations, from 2006 September 24 to 2014 March 4.
In this simultaneous fitting procedure, $N_{\rm H}$, $\textit{kT}_{\rm{cold}}$ and the cold temperature normalization are constrained to be equal at each epoch, and the other model parameters are varied independently at each epoch until a total chi-square minimum value is found.
The cold blackbody component is relatively dominant over the warm component from 2006 September 24 to 2014 March 4, when we are in the two blackbody regime of the entire three-to-two-blackbody model.
Therefore, at these later times, we obtained better (than at earlier times) constraints on the cold blackbody component normalization and temperature $\textit{kT}_{\rm{cold}}$, as well as the column density $N_{\rm H}$.
Consequently, these are the values that we chose to hold constant throughout the three-to-two-blackbody model.
We used the XSPEC model 'wabs', which uses abundances from \cite{mor83}, to calculate the effects of photoelectric absorption. 
All spectral modeling was done in the 0.3 to 10 keV energy range.
The best fit value of the column density is $N_{\rm H}=(0.945\pm0.045)\times10^{22}$ cm$^{-2}$.
This is significantly larger than the of value of $N_{\rm H}=(0.72\pm0.02)\times10^{22}$ cm$^{-2}$ found by \cite{ber09}.
This discrepancy is due to the additional new data that has now been modelled by the two blackbody regime of the three-to-two-blackbody model.
\cite{ber09} also found $kT_{\rm{cold}}=0.144\pm0.003$ keV, which is statistically consistent with our measurement of $kT_{\rm{cold}}=0.138\pm0.006$ keV.
For this two blackbody model simultaneous fit we obtained a reduced chi-square value of 1.15 for 804 degrees of freedom. 

Next we simultaneously fit the first seven \xmm \ observations, from 2003 September 8 to  2006 March 12.
For this three blackbody model, we obtained a reduced chi-square value of 1.16 for 775 degrees of freedom.  
An absorption feature at 1.2 keV was included in all of the fits described here.  
Below we describe the significance of this feature and the details of how it was incorporated into the three-to-two-blackbody model.

\begin{figure}
\includegraphics[width=0.85\linewidth]{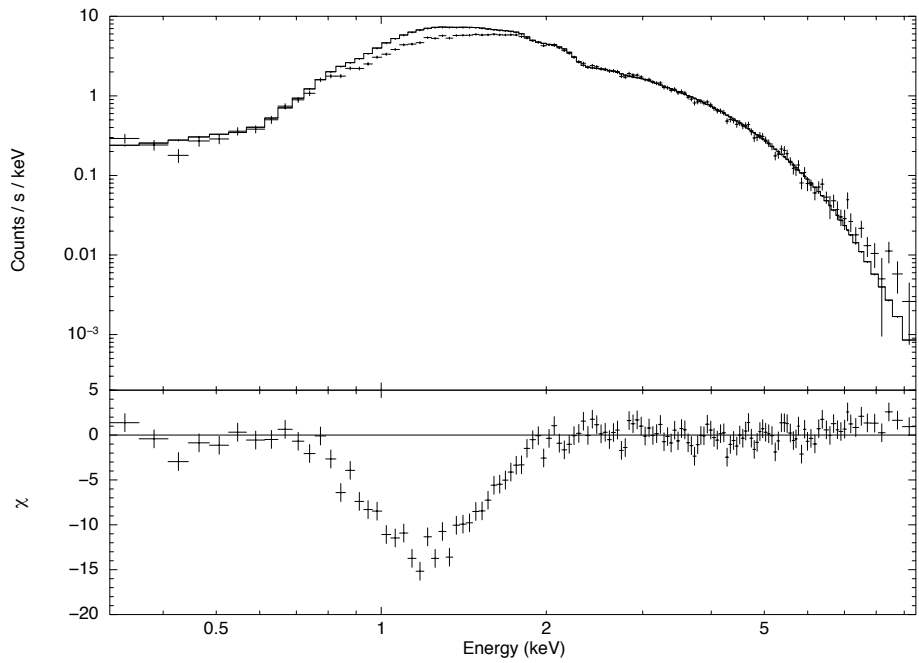}
\caption{\label{fig:line_res}
Absorption line residuals in the 2003 September 8 observation, illustrating the presence of this feature even from the beginning of \xte's outburst.
All the model parameters are held at the best fit values of Table 2, but the line strength is set to zero.   }
\end{figure}

\begin{figure}
\includegraphics[width=0.85\linewidth]{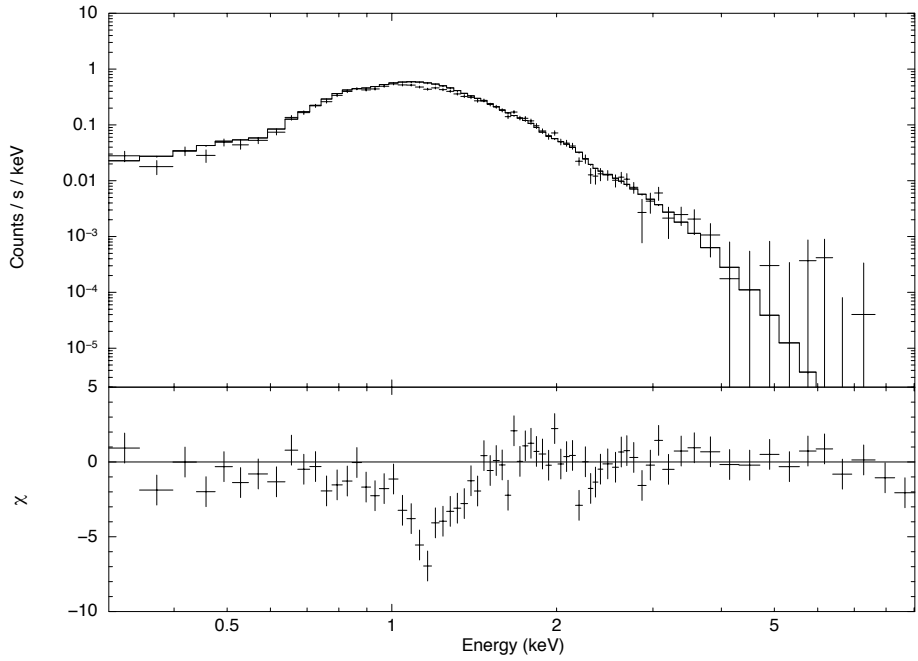}
\caption{\label{fig:11_resid}
Absorption line residuals in the 2009 March 5 observation, illustrating that the line width has decreased since 2003 September 8.
All the model parameters are held at the best fit values of Table 2, but the line strength is set to zero. 
  }
\end{figure}

\begin{figure}
\includegraphics[width=0.85\linewidth]{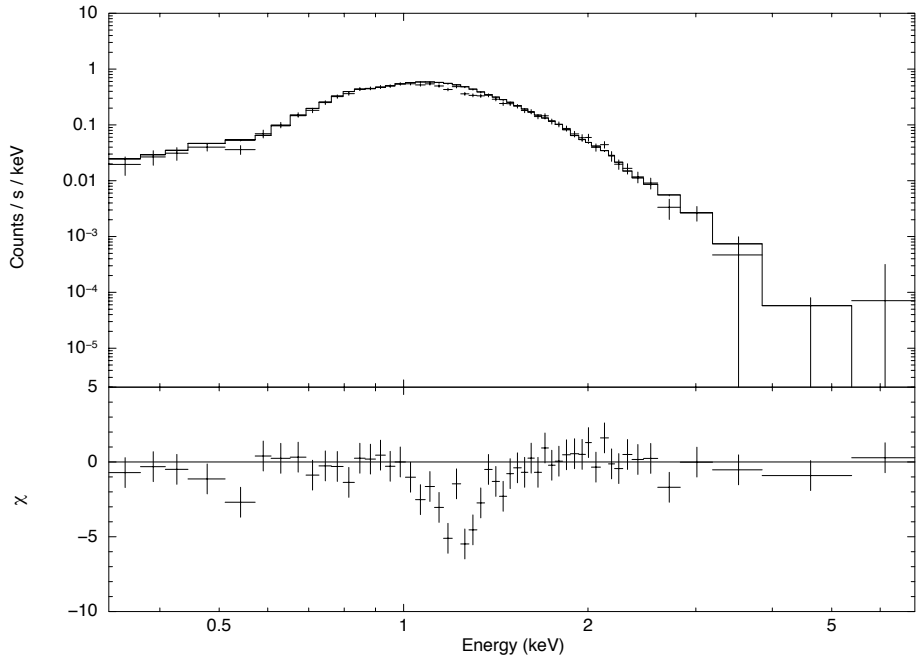}
\caption{\label{fig:22_resid}
Absorption line residuals in the most recent 2014 March 4 observation, illustrating that the absorption line has not changed since 2009 March 5.
All the model parameters are held at the best fit values of Table 2, but the line strength is set to zero. 
}
\end{figure}

The absorption feature at 1.2 keV was modeled with a Gaussian absorption line with the optical depth profile$$\tau(E)=\frac{d}{\sigma\sqrt{2\pi}}\ e^{-\frac{1}{2}\left(\frac{E-E_{\rm cyc}}{\sigma}\right)^2}.$$
This feature was also modeled by \cite{ber09}, and is likely produced by resonant proton scattering. 
If resonant proton scattering is the origin of this spectral feature, then it provides a measurement of the magnetic field in which the scattering occurs.
The energy of such an absorption line is: $E_{\rm{cyc}} = 0.63(1+z)^{-1}(B/10^{14}G)$ keV where $(1+z)^{-1}= (1-2GM/Rc^{2})^{1/2}$ is the gravitational redshift.
From 2006 September 24 to 2014 March 4, we did not find significant changes in the absorption line among the different epochs.
We therefore held the parameters of this absorption line fixed in all of the later observations, from 2006 September 24 to 2014 March 4.
We also did not find significant changes in the absorption line among the different epochs from 2003 September 8 to  2006 March 12.    
Since the line energy centroid did not differ with any statistical significance between earlier and later observations, we adopted a constant value of 1.2 keV across all epochs.
In summary, the line energy was held fixed at 1.2 keV at all epochs, but the line strength and width were allowed to differ between the two and three blackbody regimes. 
The line width $\sigma$ was about 320 eV in the three blackbody regime, and about 130 eV in the two blackbody regime.
The strength of the line $d$ decreased from 0.28 in earlier observations to 0.07 in later observations.
We also searched for evidence of phase-variability of the absorption feature by fitting the peaks and troughs of the pulse profiles separately.  
We did not find any statistically significant differences in the best fit Gaussian absorption feature parameters.
We also note that Bernardini et al. did not find it necessary to include an absorption line in the first three observations on 2003 September 8, 2003 October 12, and 2004 March 11.
However, with our higher adopted column density value, we found that an absorption feature was necessary in these observations as well.
To illustrate this, we show the line residuals of the 2003 September 8 observation in Figure \ref{fig:line_res}, where all the model parameters are held at the best fit values of Table 2, but the line strength is set to zero.
Similarly, in Figures \ref{fig:11_resid} and \ref{fig:22_resid} we show the absorption feature in the 2009 March 5 and 2014 March 4 observations. 

\begin{figure}
\includegraphics[width=1.1\linewidth]{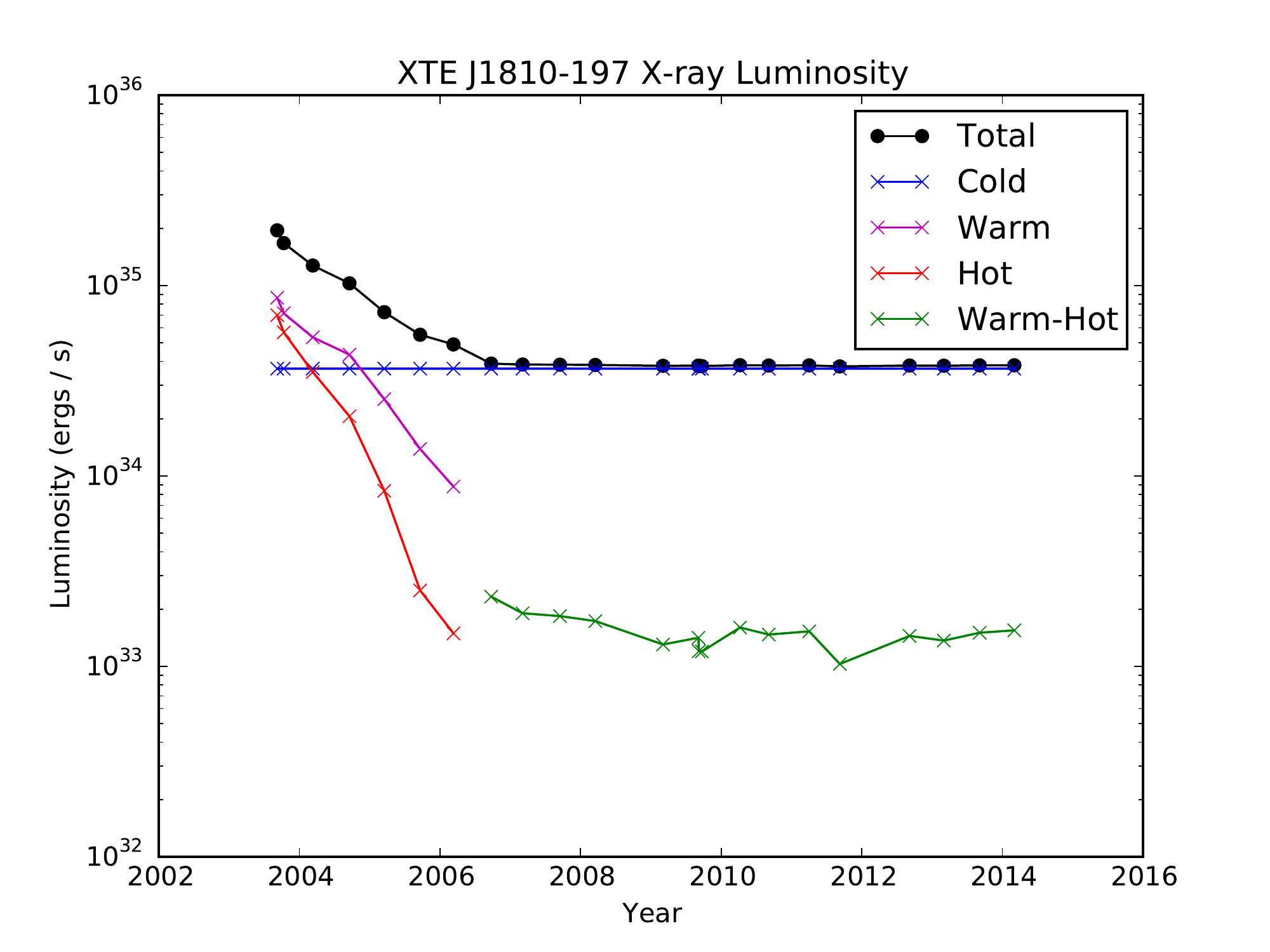}
\caption{\label{fig:xraylum}
X-ray luminosity of the components of the three-to-two-blackbody model.  
These are bolometric luminosities calculated from the temperatures and apparent areas in Table 2, assuming a distance of 3.5 kpc.}
\end{figure}

\begin{figure}
\includegraphics[width=1.1\linewidth]{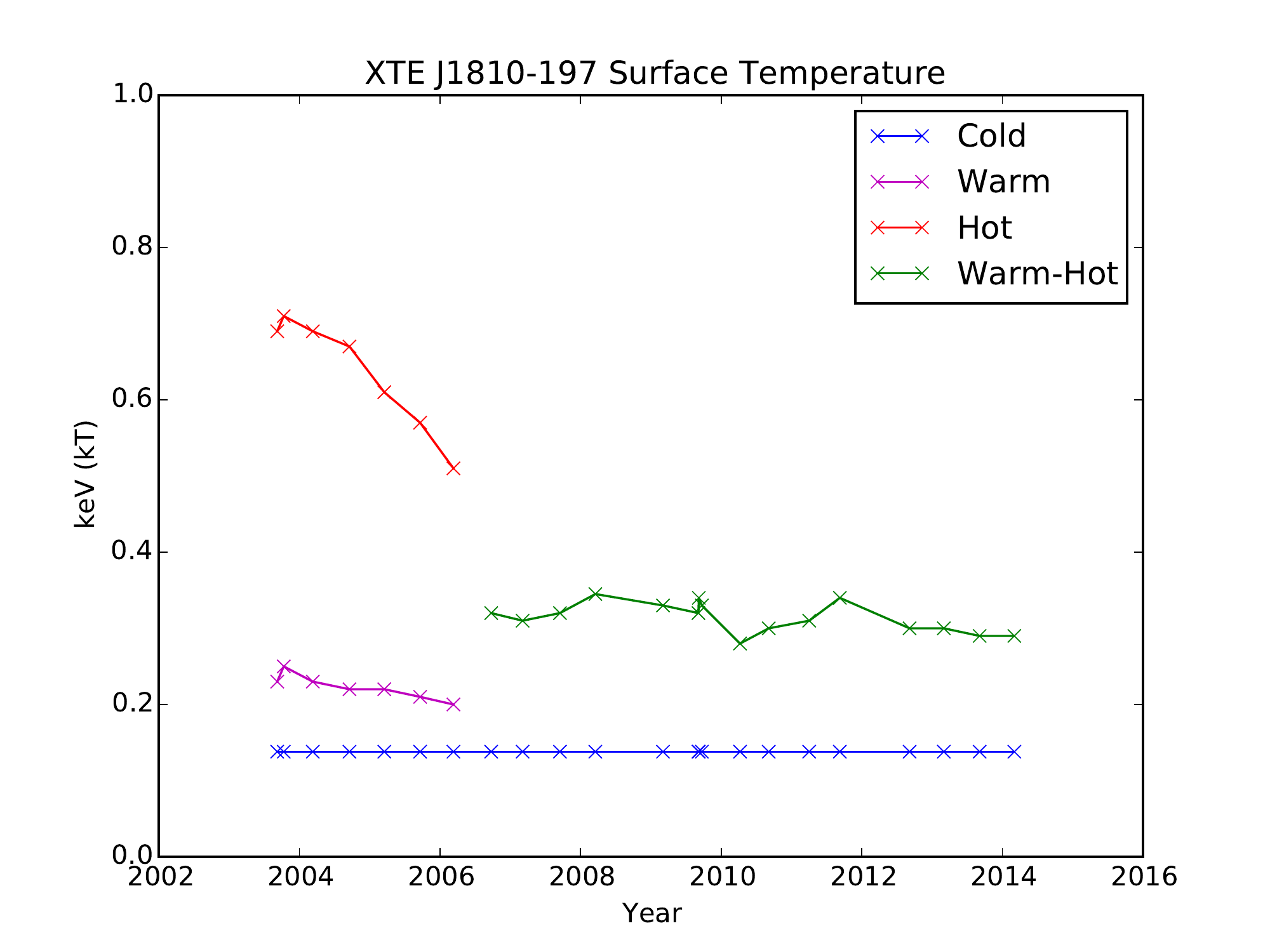}
\caption{\label{fig:temps}
Temperatures of the components of the three-to-two-blackbody model.}
\end{figure}

\begin{figure}
\includegraphics[width=1.1\linewidth]{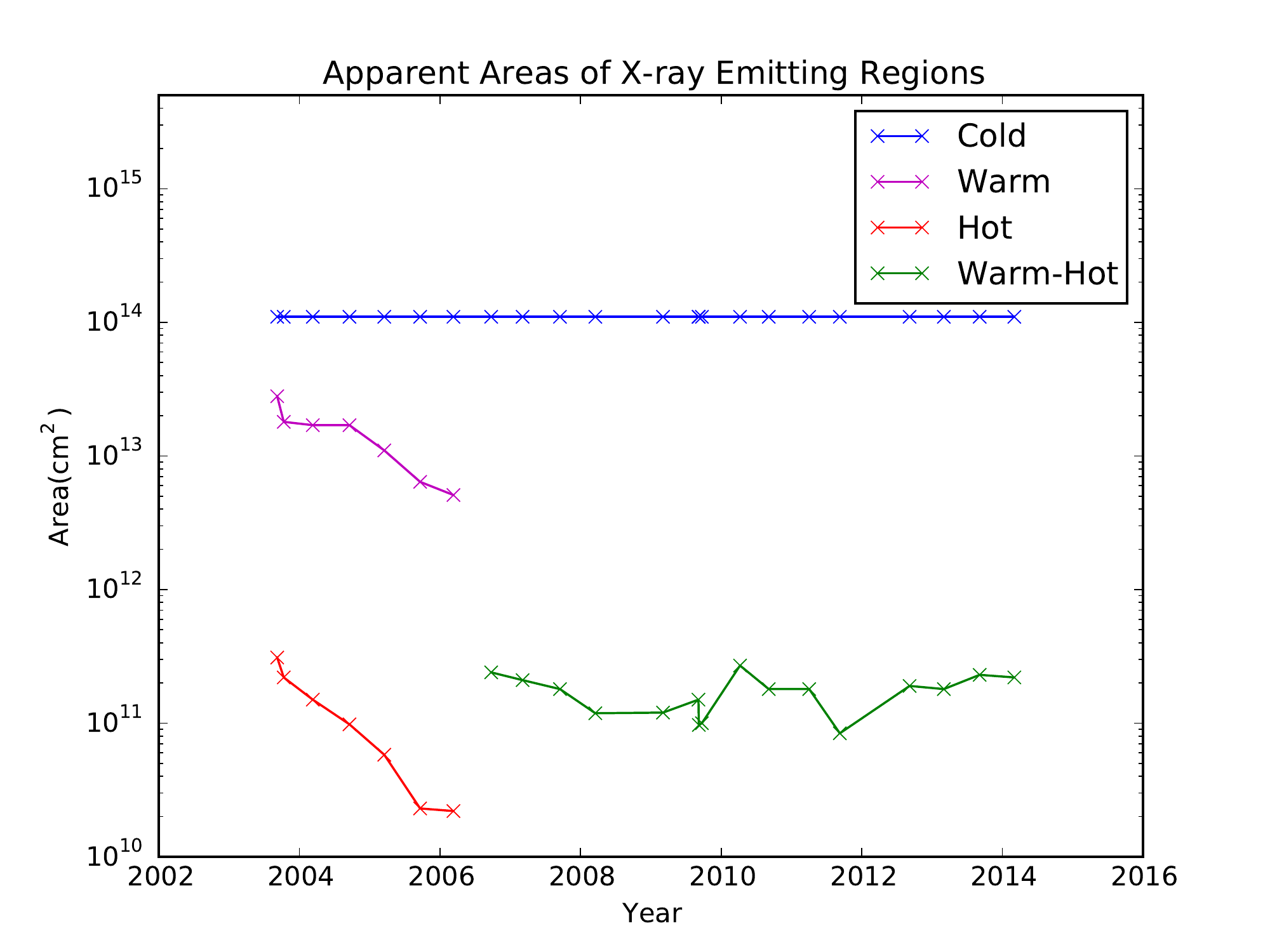}
\caption{\label{fig:areas}
Areas of the components of the three-to-two-blackbody model, assuming
a distance of 3.5 kpc.
}
\end{figure}

Table 2 lists the results of the fits to the three-to-two-blackbody model.  
Figures \ref{fig:xraylum}, \ref{fig:temps} and \ref{fig:areas} illustrate the evolution of the bolometric luminosity, temperatures and apparent emitting areas.
Because the absorption feature at 1.2 keV was characterized by a different strength and width in the two and three blackbody regimes, we set the strength of this component to zero when computing the fluxes.
This convention allows for a meaningful comparison of the fluxes as \xte \ moves from the three blackbody regime to the two blackbody regime. 
The model parameters in the three blackbody regime and the model parameters in the two blackbody regime are technically components of distinct models.
 For this reason we introduce new symbols $\textit{kT}_{\rm{warm-hot}}$, $\textit{F}_{\rm{warm-hot}}$, and $\textit{A}_{\rm{warm-hot}}$ in Table 2.
The three-to-two-blackbody model is an approximation to a gradual transition between these two regimes, and that is why there is a jump in the values of  $kT_{\rm{warm}}$ to the values of  $kT_{\rm{warm-hot}}$ when the model is switched.  

\begin{figure}
\includegraphics[width=0.85\linewidth]{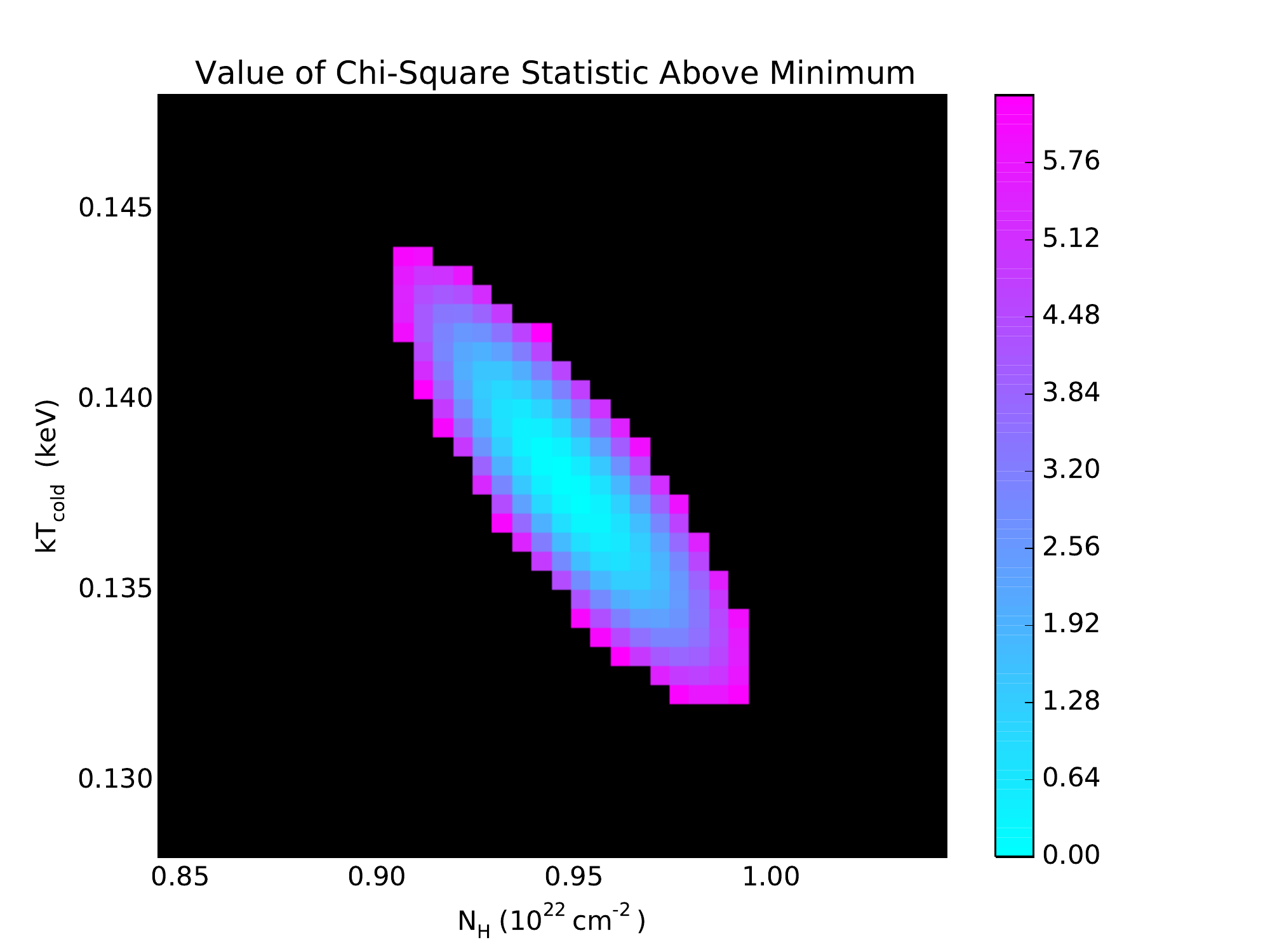}
\caption{\label{fig:chi-square}
Chi-square grid for a range of column densities and temperatures of the whole NS surface. The
region in black is outside of the 90\% confidence range for three interesting parameters.}
\end{figure}

\begin{figure}
\includegraphics[width=0.85\linewidth]{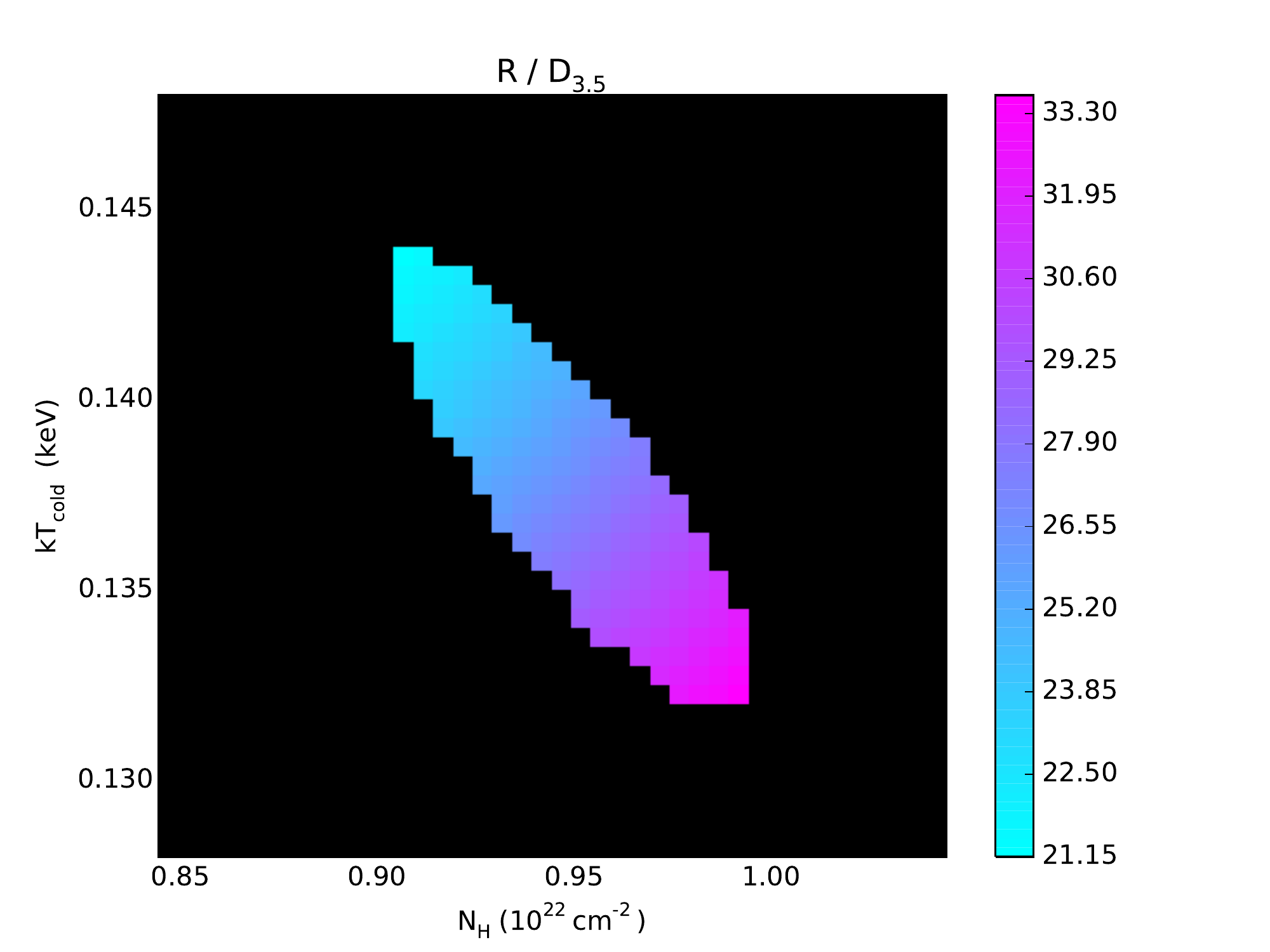}
\caption{\label{fig:radius}
Grid of NS radii \textit{R} (km) for the range of column densities and temperatures in Figure \ref{fig:chi-square}.  We assume a distance of 3.5
kpc and a mass of $1.4\,M_{\odot}$. $D_{3.5}$ is the distance to \xte \ in units of 3.5 kpc.  We account for the general relativistic correction between apparent and physical NS radius.}
\end{figure}

With the most recent data putting more constraints on the cold blackbody component, thought to represent the whole surface of the NS, we tested this whole surface interpretation for consistency with a physically plausible value of the NS radius. 
 Figure \ref{fig:chi-square} shows the chi-square statistic on a grid of values of surface temperature and column density.  
 Since the NS surface temperature parameter is most correlated with the column density, this grid provides a good estimate of the uncertainty in the NS surface temperature, and therefore also the NS radius.  
 Since none of the other model parameters (e.g. the cold component normalization) were varied, this is an underestimate of the full NS radius confidence range.  
 Figure \ref{fig:radius}  shows the inferred NS radius for all of the grid points at the top of the figure. 
 These radii values were calculated as follows:
 At each grid point the free model parameters were re-fit, and the new best fit blackbody normalization was used to calculate an apparent radius assuming a distance of 3.5 kpc.
A distance of $3.5\pm0.5$~kpc was measured by \cite{min08}.
 We then assumed a mass of 1.4 M$_{\odot}$ and applied the relativistic correction, $R_{\infty}=R\,(1-r_{\rm g}/R)^{-1/2}$, to compute the physical NS radius.
 $R_{\infty}$ is the apparent NS radius at infinity and $r_g = 2GM/c^2$ is the Schwarzschild radius.  
 We find $R_{\infty}=28.9^{+6.7}_{-5.3}$ km and $R=26.6^{+6.7}_{-5.5}$ km.
 
 This is significantly different from the physically plausible range of about $9-13$~km, and we emphasize that we are not claiming that the best fit value $R$ quoted above is a measurement of the NS radius.
 Rather, we interpret this large best fit radius as evidence that the whole surface of \xte \ is visible, and that the spectrum has some deviations from two pure blackbodies plus a Gaussian absorption feature.
 In addition to deviations from simple, uniform temperature blackbodies, we list here several factors that may contribute to this discrepancy between the physical NS radius and the best fit value. 
First, the best fit radius $R$ is highly sensitive to the shape and energy of the absorption feature.
For example, fixing the line energy at 1.3 keV decreases $R$ by 22\% while increasing the chi-square statistic to only 1.27.
 Second, the physical NS radius measurement is proportional to the NS distance, and it possible that \xte \ is at a distance closer to the low end of the range of 3.5 $\pm$ 0.5 kpc measured by \cite{min08}.
 Third, the physical NS radius measurement  decreases as the mass of \xte \ increases, and it is possible that \xte \ is more massive than the canonical value of $1.4\,M_{\odot}$. 
 NSs with masses up to $2.0\,M_{\odot}$ have been observed \citep{dem10}.
 A magnetar mass has never been measured, and it might not be surprising if magnetars as a class are toward the upper end of the range of possible NS masses \citep{mer15}. 
 Fourth, as stated above, our estimate of the confidence range of this physical NS radius measurement is actually an underestimate of the true statistical uncertainty.
 
We also checked the 2008 March 18 Chandra observation for consistency with the two blackbody model.  
This observation is notable because it was taken just months before \ \xte \ became radio quiet in late 2008 \citep{cam15}.  
We held the hydrogen column density, absorption line, and cool temperature component fixed at the \xmm \ fit values and allowed the warm temperature component to vary.    
We find that it is consistent with the rest of the later observations with a reduced chi-square fit statistic of 1.05 for 112 degrees of freedom, and a warm component temperature of 0.35 keV.

\begin{deluxetable}{l c c c c c c c c c c} 
\tabletypesize{\tiny}
\setlength{\tabcolsep}{0.009in} 
\tablecaption{Three-To-Two-Blackbody Model}
\tablehead{ \colhead{Date}  & \colhead{$\textit{kT}_{\rm{hot}}$} & \colhead{$\textit{kT}_{\rm{warm}}$/}   & \colhead{$\textit{kT}_{\rm{cold}}$} & \colhead{$\textit{F}_{\rm{hot}}$}
 & \colhead{$\textit{F}_{\rm{warm}}$/} & \colhead{$\textit{F}_{\rm{cold}}$} & \colhead{$\textit{A}_{\rm{hot}}$} & \colhead{$\textit{A}_{\rm{warm}}$/} & \colhead{$\textit{A}_{\rm{cold}}$} & \colhead{$\textit{L}_{\rm{bol}}$}   \\
 && \colhead{$\textit{kT}_{\rm{warm-hot}}$} &&& \colhead{$\textit{F}_{\rm{warm-hot}}$ } &&& \colhead{$\textit{A}_{\rm{warm-hot}}$} &&
 \\
 \colhead{(UT)}  & \colhead{(keV)} & \colhead{(keV)} & \colhead{(keV)} & \colhead{(erg cm$^{-2}$ s$^{-1}$)}
 & \colhead{(erg cm$^{-2}$ s$^{-1}$)} & \colhead{(erg cm$^{-2}$ s$^{-1}$)} & \colhead{(cm$^2$)} & \colhead{(cm$^2$)} & \colhead{(cm$^2$)}  & \colhead{($10^{34}$erg s$^{-1}$)}   % $10^{34}$  erg cm$^{-2}$ s$^{-1}$
 }
\startdata
2003/9/8 & 0.69$\pm$0.01 & 0.23$\pm$0.01 & 0.138$\pm$0.006 & $3.2 \times 10^{-11}$ & $8.1\times 10^{-12}$ &  $5.8 \times 10^{-13}$ & $3.1 \times 10^{11}$ &  $2.8\times10^{13}$ & $1.1\times 10^{14}$ & 19.5  \\
2003/10/12  & 0.71$\pm$0.02 & 0.25$\pm$0.01 & $^{\prime\prime}$ & $2.6\times 10^{-11}$ & $6.7 \times10^{-12}$ & $^{\prime\prime}$  & $2.2\times10^{11}$ & $1.8\times10^{13}$ & $^{\prime\prime}$ & 16.7 \\
2004/3/11  & 0.69$\pm$0.02 & 0.23$\pm$0.01 & $^{\prime\prime}$  & $1.6\times 10^{-11}$ & $4.2 \times10^{-12}$ & $^{\prime\prime}$  & $1.5\times10^{11}$ & $1.7\times10^{13}$ & $^{\prime\prime}$  & 12.7  \\                   
2004/9/18    & 0.67$\pm$0.01  & 0.22$\pm$0.01 & $^{\prime\prime}$  &$9.0\times 10^{-12}$ & $3.0 \times10^{-12}$ & $^{\prime\prime}$ & $9.8\times 10^{10}$ &  $1.7 \times 10^{13}$ & $^{\prime\prime} $ & 10.3 \\
2005/3/18    & 0.61$\pm$0.01 & 0.22$\pm$0.01  &  $^{\prime\prime}$ & $3.4\times 10^{-12}$ & $1.6 \times10^{-12}$ & $^{\prime\prime}$ & $5.8\times 10^{10}$ & $1.1 \times 10^{13}$  & $^{\prime\prime}$ & 7.3   \\
2005/9/20   & 0.57$\pm$0.03 & 0.21$\pm$0.01  & $^{\prime\prime}$ & $9.5\times 10^{-13}$  & $8.4 \times 10^{-13}$ & $^{\prime\prime}$   & $2.3\times 10^{10}$ & $6.4 \times 10^{12}$ & $^{\prime\prime}$ & 5.5\\
2006/3/12   & 0.51$\pm$0.05 & 0.20$\pm$0.02 & $^{\prime\prime}$ & $5.0\times 10^{-13}$  & $4.5 \times10^{-13}$ & $^{\prime\prime}$  & $2.2\times 10^{10}$  & $5.1 \times 10^{12}$ &$^{\prime\prime}$ & 4.9 \\
2006/9/24   & & 0.32$\pm$0.01 &  $^{\prime\prime}$ &  & $4.2   \times10^{-13}$  & $^{\prime\prime}$  &  & $2.4  \times 10^{11}$  & $^{\prime\prime}$  &  3.9 \\
2007/3/6   &  &  0.31$\pm$0.01 & $^{\prime\prime}$ &  & $3.3   \times 10^{-13}$  & $^{\prime\prime}$ &&  $2.1  \times 10^{11}$  & $^{\prime\prime}$ &  3.8 \\
2007/9/16  &  & 0.32$\pm$0.01 &  $^{\prime\prime}$ &  & $3.4   \times 10^{-13}$ & $^{\prime\prime}$  &&   $1.8 \times 10^{11}$  & $^{\prime\prime}$ &  3.8\\
2009/3/5    & & 0.33$\pm$0.02 &  $^{\prime\prime}$ &   & $2.5   \times  10^{-13}$ & $^{\prime\prime}$ & &     $1.2 \times 10^{11}$  & $^{\prime\prime}$ & 3.8  \\
2009/9/5  &   & 0.32$\pm$0.02 &  $^{\prime\prime}$ & & $2.5   \times 10^{-13}$ &$^{\prime\prime}$ &&    $1.5 \times 10^{11}$   & $^{\prime\prime}$ & 3.8  \\
2009/9/7   &  & 0.34$\pm$0.03 &  $^{\prime\prime}$ & & $2.4  \times 10^{-13}$ & $^{\prime\prime}$ & &     $9.7 \times 10^{10}$ & $^{\prime\prime}$  & 3.8 \\
2009/9/23  & & 0.33$\pm$0.03 &  $^{\prime\prime}$ &  & $2.3  \times 10^{-13}$ & $^{\prime\prime}$  &&   $1.0 \times 10^{11}$  & $^{\prime\prime}$ & 3.8\\
2010/4/9  &  & 0.28$\pm$0.05 &  $^{\prime\prime}$ &  & $2.2   \times  10^{-13}$ &$^{\prime\prime}$ &&   $2.7 \times 10^{11}$  & $^{\prime\prime}$ & 3.8  \\
2010/9/5 &  &0.30$\pm$0.03 &  $^{\prime\prime}$ &  & $2.4   \times  10^{-13}$ &$^{\prime\prime}$  &&  $1.8 \times 10^{11 }$ & $^{\prime\prime}$ &   3.8 \\
2011/4/3  &  & 0.31$\pm$0.02 &  $^{\prime\prime}$ & &  $2.5   \times 10^{-13}$ & $^{\prime\prime}$ &&   $1.8 \times 10^{11}$  & $^{\prime\prime}$ &  3.8  \\
2011/9/9 &  &0.34$\pm$0.04 &  $^{\prime\prime}$ &  & $2.1  \times  10^{-13}$ & $^{\prime\prime}$ &&  $8.4 \times 10^{10}$  & $^{\prime\prime}$  & 3.8  \\
2012/9/6  &   & 0.30$\pm$0.03 & $^{\prime\prime}$ & & $2.3  \times 10^{-13}$ & $^{\prime\prime}$ &&   $1.9 \times 10^{11}$ & $^{\prime\prime}$  &  3.8  \\
2013/3/3   &  &  0.30$\pm$0.04 & $^{\prime\prime}$ & &$2.2  \times 10^{-13}$  &$^{\prime\prime}$ &&  $1.8 \times 10^{11}$   & $^{\prime\prime}$ &  3.8  \\
2013/9/5 & & 0.29$\pm$0.02  &  $^{\prime\prime}$ & & $2.2   \times 10^{-13}$ & $^{\prime\prime}$ &&   $2.3  \times 10^{11}$ & $^{\prime\prime}$ &  3.8  \\
2014/3/4 &  & 0.29$\pm$0.02 & $^{\prime\prime}$ & & $2.3  \times 10^{-13}$ & $^{\prime\prime}$ &&   $2.2 \times 10^{11}$ & $^{\prime\prime}$ &   3.8  \\
\enddata
\tablecomments{
The column density was held fixed at $N_{\rm H}=0.945\times10^{22}$~cm$^{-2}$ for all observations. 
All listed fluxes are the absorbed values and are computed with the strength of the 1.2 keV absorption feature set to zero.
Uncertainties in the hot and warm temperatures are 90\% for two interesting parameters.  
They were estimated from $\chi^2$ contours with $kT_{\rm cold}$.  
}
\end{deluxetable}

\subsection{Two Blackbody Model}

We also simultaneously fit a two blackbody model to the first seven observations, as originally considered by \cite{got07}. 
 In this model the two temperatures are thought of as a central hot spot on the surface of the NS, surrounded by a warm temperature annulus.  
We also include a Gaussian absorption line at 1.2 keV.   
We again found this model to be a good fit to the first seven observations, although with lower values of the hydrogen column density than in the three-to-two-blackbody model.  
All of these results are consistent with the findings of \cite{got07}.  
We found a reduced chi-square of 1.15 for 775 degrees of freedom.  
In light of the most recent data indicating a larger column density than previously measured, we also attempted to fit a two blackbody model with a column density fixed at this new larger value to the earlier observations.  
This resulted in a slightly worse fit with a reduced chi-square of 1.24 for 776 degrees of freedom.  
Table 3 lists the results of the fits to the two blackbody model. 

We then tried to fit the this same two blackbody model, (i.e. with the  column density value fixed at $N_{\rm H}= 0.76 \times10^{22}$ cm$^{-2}$) to the latest observations and found it was a poor fit to the data with a reduced chi-square value of 1.5.  
The lower value of the column density resulted in larger temperatures for the cool blackbody component, averaging around 0.16 keV.
This poor fit leads us to favor the larger $N_{\rm H}$ value and the three-to-two-blackbody model.

\begin{deluxetable}{l c c c c c c c}
\tabletypesize{\tiny}
\tablecaption{Two Blackbody Model}
\tablehead{  \colhead{Date} & \colhead{$\textit{kT}_{1}$} & \colhead{$\textit{kT}_{2}$} & \colhead{$\textit{F}_{1}$} & \colhead{$\textit{F}_{2}$} & \colhead{$\textit{A}_{1}$} & \colhead{$\textit{A}_{2}$} & \colhead{\textit{L}$_{\rm bol}$}   \\
\colhead{(UT)} & \colhead{(keV)} & \colhead{(keV)} & \colhead{(erg cm$^{-2}$ s$^{-1}$)} & \colhead{(erg cm$^{-2}$ s$^{-1}$)} & \colhead{(cm$^2$)}  & \colhead{(cm$^2$)} & \colhead{(erg s$^{-1}$)}  
}
\startdata
2003 Sep 8   &  0.70  $\pm$ 0.01  & 0.25 $\pm$ 0.01 & 3$.3 \times 10^{-11}$ & $7.6 \times  10^{-12}$  &  $1.2 \times 10^{13}$ & $2.7 \times 10^{11}$  & $1.2 \times 10^{35}$\\
2003 Oct 12   & 0.72   $\pm$ 0.02 & 0.27 $\pm$ 0.02 &  $2.6 \times 10^{-11}$ & $6.7 \times  10^{-12}$   & $8.4 \times 10^{12}$  & $2.0 \times 10^{11}$ & $1.0 \times 10^{35}$ \\
2004 Mar 11  & 0.70  $\pm$ 0.02 & 0.25  $\pm$ 0.02 &  $1.6 \times 10^{-11}$ & $4.3  \times 10^{-12}$   & $9.8 \times 10^{12}$ & $1.4 \times 10^{11}$  &  $7.1 \times 10^{34}$ \\                   
2004 Sep 18   & 0.68 $\pm$  0.01 & 0.23 $\pm$ 0.01 &  $9.3 \times 10^{-12}$ &  $3.2 \times  10^{-12}$  & $1.1 \times 10^{13}$ & $9.1 \times 10^{10}$ &  $5.2 \times 10^{34}$ \\
2005 Mar 18   & 0.61 $\pm$  0.01 & 0.21 $\pm$ 0.01 &  $3.6 \times 10^{-12}$ &  $1.9 \times  10^{-12}$  & $1.1 \times 10^{13}$ & $5.8  \times 10^{10}$   & $3.2 \times 10^{34}$  \\
2005 Sep 20   & 0.55 $\pm$  0.02& 0.19 $\pm$ 0.01 &  $1.1 \times  10^{-12}$ & $1.2 \times  10^{-12}$   & $1.4 \times 10^{13}$ & $2.8 \times 10^{10}$  &  $2.2 \times 10^{34}$ \\
2006 Mar 12   & 0.50 $\pm$ 0.04 & 0.18 $\pm$ 0.01 & $5.5 \times  10^{-13}$  &  $8.4 \times  10^{-13}$  & $1.7 \times 10^{13}$ & $2.4 \times 10^{10}$   & $1.9 \times 10^{34}$ \\
\enddata
\tablecomments{
The column density was held fixed at $N_{\rm H}=0.76\times10^{22}$~ cm$^{-2}$ for all observations. 
All listed fluxes are the absorbed values and are computed with the strength of the 1.2 keV absorption feature set to zero.
Uncertainties in the hot and warm temperatures are 90\% for two interesting parameters.  They were estimated from $\chi^2$ contours of the hot versus warm temperature parameters.
}
\end{deluxetable}

\subsection{Comptonized Blackbody Model}

We explored the possibility that the deviation from a single blackbody spectrum is simply due to Compton scattering.  
We attempted to fit a model where the source photons are Comptonized by relativistic electrons of small optical depth (as described in Rybicki and Lightman 1979, section 7.5)  to the first seven observations.
The blackbody spectrum is comptonized such that it is characterized by the parameter $\alpha = -{\ln \tau_{es}}/{\ln A}$ where $\tau_{es}$ is the optical depth and $A$ is the mean energy amplification per scattering. 
We allowed $\alpha$ to vary between observations, and fit each observation separately.
The fit to each individual observation was poor.
For example, the 2003 September 8 observation was one of the data sets best described by this model and had a reduced chi-square of 1.9 for 133 degrees of freedom.
We interpret this as evidence that the deviation from a simple blackbody spectrum is not dominated by Compton scattering.

\subsection{Model-Independent Measurements of Spectral and Temporal Changes}

\begin{figure}
\centerline{\includegraphics[width=1.0\linewidth]{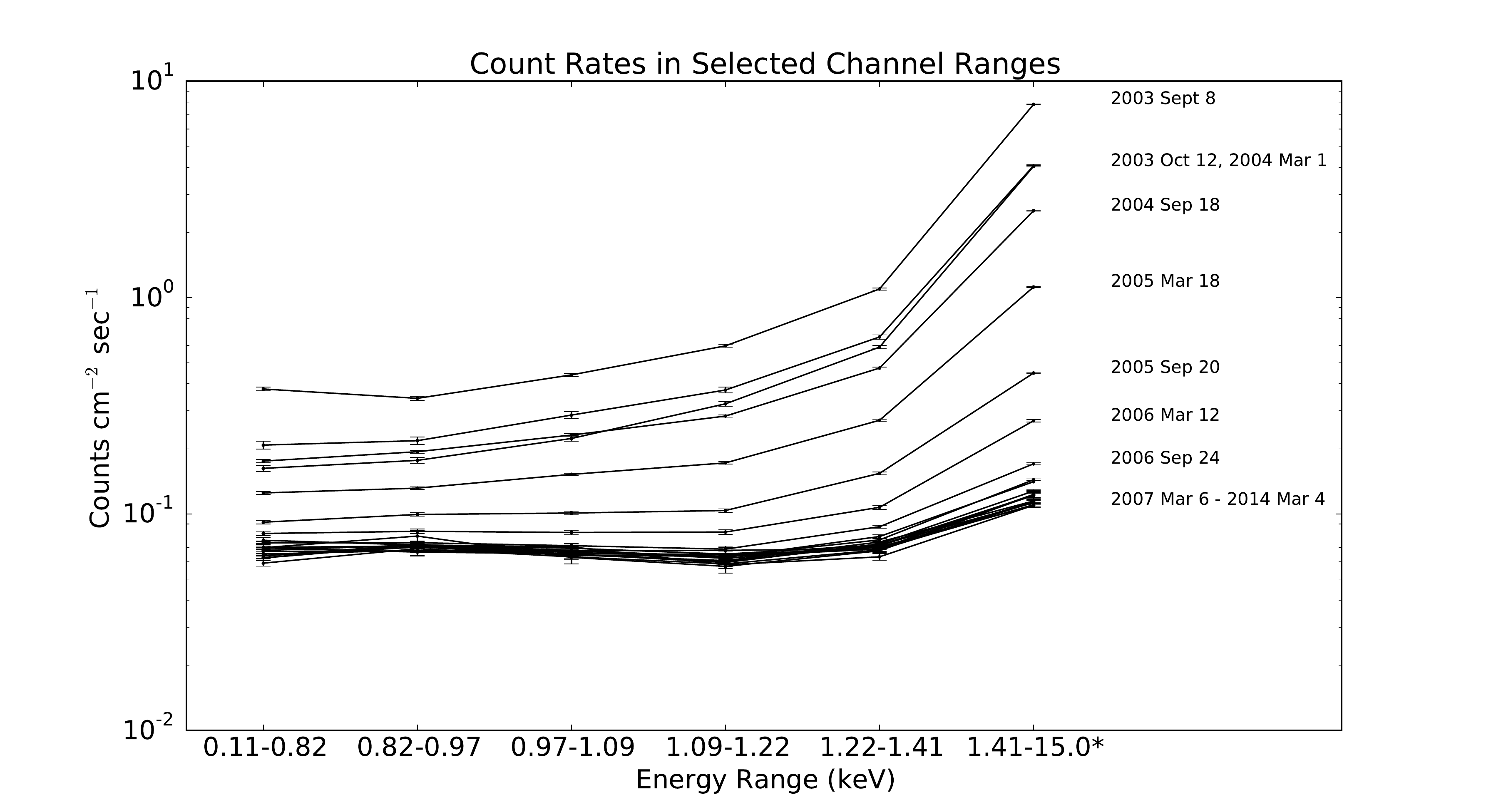}}
\caption{\label{fig:ratios}
Changes in count rate from one observation to the next. \xte's X-ray spectrum has reached a steady state.
15.0* keV is a nominal upper limit.  Most events detected in this range are at much lower energies.}
\end{figure}

We also sought to confirm, in a model independent way, that \xte \ has reached a steady state.
In Figure \ref{fig:ratios}, we plot the count rates of successive observations, for all of the \xmm \ data.
The channel ranges were chosen to keep at least 500 counts per bin.
Due to the long term stability of the EPIC pn detector, we are confident that all these channels correspond to the energy ranges shown.
We are able to confirm that \xte's X-ray spectrum has reached a steady state.

\begin{figure}
\centerline{\includegraphics[height=1.0\textheight,keepaspectratio]{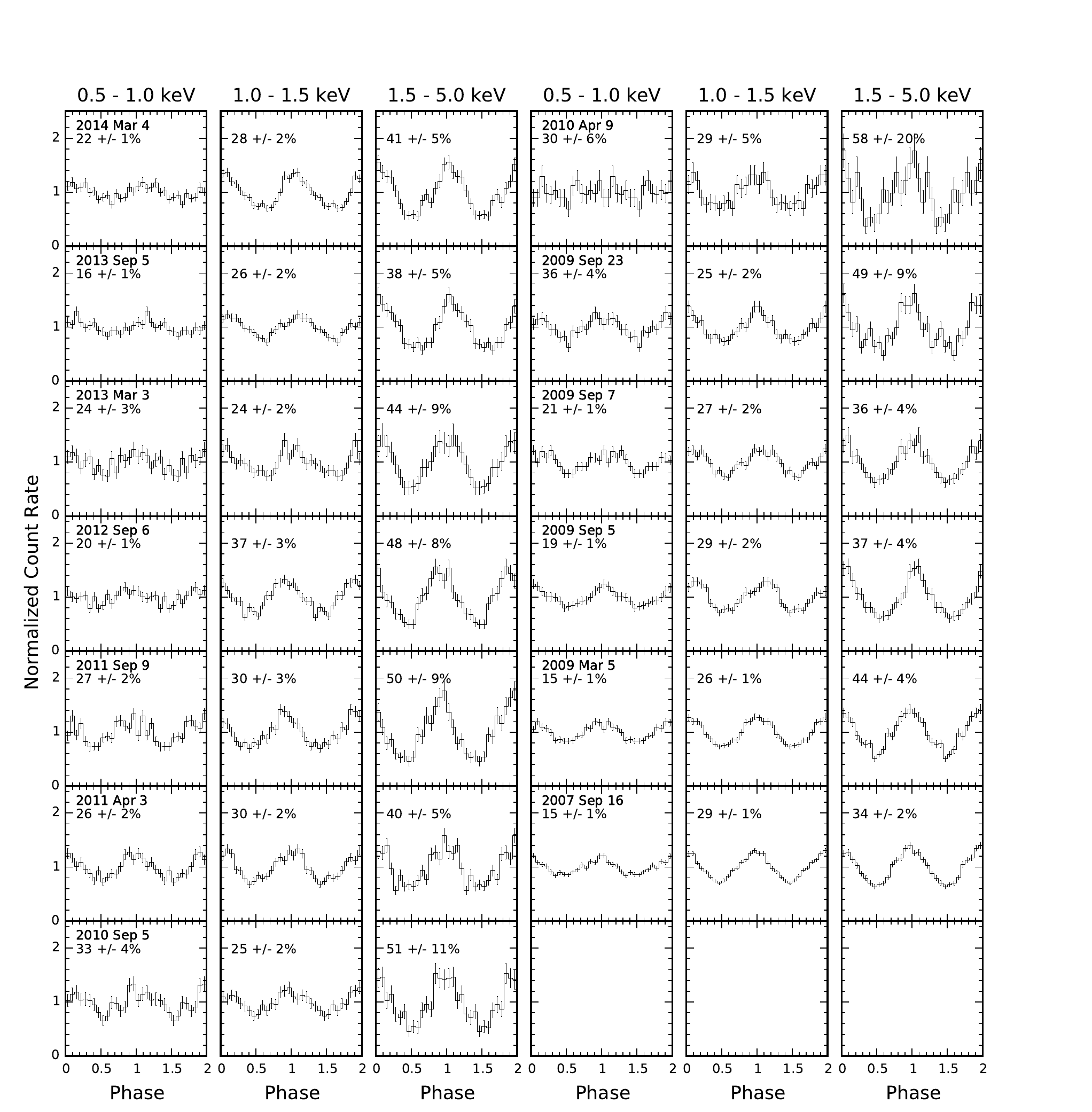}}
\caption{\label{fig:pp}
Background subtracted normalized pulse profiles. Pulsed X-ray emission from a hot region on the NS surface is evidence of continued magnetar activity. Pulsed fraction is indicated in each panel. Profiles were phase shifted to alignment in the $1.5-5.0$ keV energy range.}
\end{figure}

Another model independent test of the X-ray stability of \xte \ is its energy dependent pulse profiles, which are shown in Figure \ref{fig:pp}.
The timing analysis used to compute these pulse profiles is presented \cite{cam15}.
These pulse profiles consistently show more pulsed emission at higher energies, which is consistent with our model of the pulsed emission coming from a small warm spot on the NS surface.

\section{Discussion}
\subsection{Comparison with Previous Results}

Our findings on the previously analyzed data are consistent with the results of \cite{got07} and \cite{ber09}.  
The lower total X-ray fluxes of the most recent observations allow us to better constrain the properties of the cooler blackbody component as well at the hydrogen column density.  
We measure the column density to be about $50\%$ larger than previously reported.
There is an abundance of new X-ray data in this paper that was not available to \cite{got07} and \cite{ber09}.
It is this new data that has led to the higher measured column density.
We also note that the best fit whole surface temperature we have measured, 0.138 keV, is statistically consistent (within the 90\% uncertainty range) with the value in Bernardini et al. 2009 (0.144 keV).

The NS radius value quoted in Bernardini et al. is $R_{\infty}= 17.9\pm^{+1.9}_{-1.5}$ km and assumes a distance of 3.5 kpc.  
This is significantly smaller than our measurements of $R_{\infty}=28.9^{+6.7}_{-5.3}$ km and $R=26.6^{+6.7}_{-5.5}$ km.
This discrepancy is explained by the larger column density that is required by the new X-ray data we have from 2009 March 5 to 2014 March 4.

\subsection{Has \xte\ ``Returned to Quiescence''?}

Between the time of the 2008 March 18 \chandra \ observation and the 2009 March 5 \xmm \ observation \xte \ became radio quiet \citep{cam15}.
It is notable that the last significant ($\approx 20\%$) decrease in the warm/hot flux occurred in this same time interval. 
This suggests a possible correlation between this magnetar's X-ray hot spot flux and its radio emission.   
We find no other correlation between \xte's \  radio turn off and its X-ray behavior during the \chandra \ observation.
 
Beginning with the 2009 March 5 observation, the total X-ray flux  %(in the .3 to 10 KeV range) 
of \ \xte \  reached a constant minimum value. %  of about 8E-13 erg $cm^{-2}s^{-1}$. 
The warm temperature component is also constant to within statistical uncertainties. 
However, it is not clear that \ \xte \ is back in its pre-outburst state, i.e. that it has returned to quiescence.  
The archival \textit{ROSAT} data is not of high enough quality to tightly constrain its previous surface temperature or the possible existence of an absorption feature.  
Furthermore, Bernardini et al. (2009) found that both one and two blackbody models are good fits to the \textit{ROSAT} data.  
Gotthelf et al. (2004) fit a single blackbody model to four archival \textit{ROSAT} observations, spanning 1990 September 3 to 1993 April 3, and calculated unabsorbed X-ray fluxes 
ranging from $(5.5-8.3) \times 10^{-13}$ erg cm$^{-2}$ s$^{-1}$ (in the $0.5-10$ keV range). 
From 2009 March 5 to 2014 March 4 we measure unabsorbed X-ray fluxes ranging from $(7.4-7.8) \times 10^{-13}$  erg cm$^{-2}$ s$^{-1}$ in the $0.5-10$ keV range, in accord with the \textit{ROSAT} measurements. 

This evidence suggests that it is possible that \xte \ is back in its pre-outburst, ``quiescent'' state.
However, given the uncertainty of the details of its pre-outburst state, it is just as possible that \xte's magnetosphere is in a new configuration.
The hot spot on its surface and the 1.2 keV absorption line may even be new features.

\subsection{Comparison with Other Transient Magnetars}

Several other transient magnetars have been found since the discovery of \xte \ in 2003.
Here we review what is known about the later stages of these other transient magnetars.
First we note that there is no other transient magnetar whose whole surface temperature has been measured.
This is likely due to a combination of relatively large distances to and low surface temperatures of the other transient magnetars.
We also note that none of the blackbody components of the transient magnetars have been observed to completely fade away after an outburst.
(A caveat is that 3XMM J185246.6+003317 was only observed for several months after its 2008 outburst, and its current state is therefore unknown.)

Table 4 lists some of the properties of the known transient magnetars as well as the candidate transient magnetars PSR J1622$-$4950 and AX J1845$-$0258.
We list the most recently measured spin parameters as well as the most recently measured X-ray luminosity.
In the interest of comparing their current spin-down luminosities $\dot E$ to current bolometric X-ray luminosities, we quote the X-ray luminosities in the widest bands reported.

In addition to \xte, there are several other transient magnetars whose X-ray luminosity is likely magnetically powered several years post outburst.
They are the "low magnetic field" magnetar SGR 0418+5729, SGR 0501+4516, CXO 164710.2$-$455216, the Galactic center magnetar SGR J1745$-$2900, and Swift J1822.3$-$1606.
The Galactic center magnetar SGR J1745$-$2900 is particularly similar to \xte \ in that it produced radio emission and had a relatively slow decay as its blackbody temperature and hot spot area decreased.

There are also three transient magnetars whose spin-down power exceeds their X-ray luminosity.
They are 1E 1547.0$-$5408, SGR 1627$-$41, and Swift J1834.9$-$0846.
1E 1547.0$-$5408 is still much more luminous than its lowest pre-outburst state, and is therefore not in a true quiescent state \citep{ber09}.
SGR 1627$-$41 and Swift J1834.9$-$0846 are currently close to their pre-outburst fluxes (an upper bound in the case of Swift J1834.9$-$0846) and could be in true quiescent states \citep{an12,kar12,esp13}.

In some cases, there are observations of transient magnetars before their outburst.
As alluded to above, a pre-outburst observation of Swift J1834.9$-$0846 with Chandra could not even detect the source, and gave an upper limit on the $2-10$~keV luminosity of $1.7 \times 10^{31}$ erg~s$^{-1}$ \citep{kar12}.
This could be a pre-outburst quiescent state in which there is no magnetar activity.
SGR 1627$-$41 was observed between its 1998 and 2008 outbursts, and its luminosity is similar several years after each outburst \citep{an12}.
A non-detection of 3XMM J185246.6+003317 by Chandra in 2001 gives an upper bound of $4 \times 10^{32}$ erg~s$^{-1}$ \citep{zho14}.
SGR 0501+4516 was observed by ROSAT before its 2008 outburst and, like \xte, had a similar temperature and flux level post-outburst \citep{cam14}.
SGR 1833$-$0832 could not be detected by \xmm \ in 2006, five years before its 2011 outburst \citep{esp11} 

While many transient magnetars were fit to blackbody plus powerlaw models, some could be fit to pure blackbody models with temperatures comparable to \xte.
The low magnetic field magnetar SGR 0418+5729 has a very small hot spot of temperature 0.3 keV \citep{rea13}.
PSR J1622$-$4950 has a hot spot temperature of about 0.5 keV \citep{and12}.
SGR 1833$-$0832 is best modeled by a 1.2 keV blackbody \citep{esp11}.
The distance to SGR 1833$-$0832 is highly uncertain, so we don't know if its X-ray luminosity or spin-down power is dominant.
The blackbody components of transient magnetars that were modeled by a blackbody plus a power-law are all within about a factor of 2 of these temperature values.

AX J1845$-$0258 is included in this table even though its status as a transient magnetar is uncertain.
It has disappeared since its 1993 outburst and its period derivative has not been measured \citep{tor98,tam06}.
The other candidate transient magnetar, PSR J1622$-$4950, was discovered in radio and subsequently observed as an X-ray source.
Its fading X-ray emission suggests that this radio magnetar could have been observed just after an outburst in 2007 \citep{and12}.

\begin{deluxetable}{l l l c c l c c}
\tabletypesize{\tiny}
\tablewidth{0pt}
\setlength{\tabcolsep}{0.04in} 
\tablecaption{Transient Magnetars}
\tablehead{  \colhead{Name} &  \colhead{Outburst Obs. Date} &  \colhead{Period} & \colhead{$\dot{P}$} & \colhead{$\dot{E}$} & \colhead{$L_{x}$} & \colhead{Radio?}
& \colhead{References}
\\
& \colhead{} & \colhead{(s)} & \colhead{($10^{-11}$ s s$^{-1}$)} & \colhead{($10^{33}$ erg s$^{-1}$)} & \colhead{(10$^{33}$ erg s$^{-1}$)}
}
\startdata 
SGR 0418+5729 & 2009 June & 9.07838822(5) & 0.0004(1) & 0.00021 & $\approx0.006$ ($0.5-10$ keV) & No &  1\\
SGR 0501+4516 & 2008 Aug & 5.7620695(1) & 0.594(2) & 1.2 & $\approx9$ ($0.5-10$ keV) &  No & 2 \\
1E 1547.0$-$5408 & 2008 Oct, 2009 Jan & 2.0721255(1) & 4.77 &  210  & $\approx25$ ($1-10$ keV) &  Yes & 3,4 \\
PSR J1622$-$4950* & ... & 4.3261(1) & 1.7(1) & 8.3 & $\approx1$ ($0.3-10$ keV) &  Yes & 5,6 \\
SGR 1627$-$41 & 1998 June, 2008 May & 2.594578(6) & 1.9(4) & 43 & $\approx$ 3 ($2-10$ keV) &  No & 7,8 \\
CXO 164710.2$-$455216 & 2006 Sep & 10.610644(17) & $<0.04$ & $<0.013$ & $\approx$ 10 ($2-10$ keV) &  No & 9 \\
SGR J1745$-$2900 & 2013 June & 3.76363824(13) & 1.385(15) & 10 & $\approx70$ ($1-10$ keV) &  Yes & 10,11 \\
\xte  & 2003 Sep & 5.540525412(3) & 2.79039(6)  & $0.56-0.66$ & $\approx38$ ($0.3-10$ keV) &  Yes & 12 \\
Swift J1822.3$-$1606 & 2011 July & 8.43772106(6) & 0.00214(21) & 0.0014 & $\approx0.1$ ($1-10$ keV) &  No & 13 \\
SGR 1833$-$0832 & 2010 Mar & 7.5654084(4) & 0.35(3)  & 0.32 & ... & No &14 \\
Swift J1834.9$-$0846 & 2011 Aug & 2.4823018(1) & 0.796(12) & 21 & $\approx0.057$ ($2-10$ keV) &  No & 15,16 \\
AX J1845$-$0258* & 1993 & 6.97127(28) & ... & ... & ... &  No & 17 \\
3XMM J185246.6+003317 & 2008 Sep & 11.55871346(6) & $<0.014$ & $<0.0036$ & $\approx3$ ($0.5-10$ keV) & No & 18 \\
\enddata
\tablecomments{
*PSR J1622$-$4950 is only a candidate transient magnetar, since no X-ray outbursts have yet been observed.  
AX J1845$-$0258 has not been observed since its 1993 outburst, and no period derivative has been measured.
}
\tablerefs{
(1) \citealt{rea13};
(2) \citealt{cam14};
(3) \citealt{dib12};
(4) \citealt{kui12}
(5) \citealt{and12}
(6) \citealt{lev10}
(7) \citealt{an12}
(8) \citealt{esp09}
(9) \citealt{an13}
(10) \citealt{cot15}
(11) \citealt{kas14}
(12) \citealt{cam15}
(13) \citealt{sch14}
(14) \citealt{esp11}
(15) \citealt{esp13}
(16) \citealt{kar12}
(17) \citealt{tor98}
(18) \citealt{rea14}
}
\end{deluxetable}

\subsection{Comparison with Theory}

\cite{bel09} predicted that the hot spot on the surface of \xte \ would fade away.  
In the special case of a narrow j-bundle with a uniform twist, Beloborodov predicts that the hot spot luminosity will decay with a timescale
 $t_{ev}\approx15\,V_{9}^{-1}\,B_{14}\,R_{6}^{2}\,\Psi\,u_{*}$ years  where $V_{9}$ is the voltage in units of 10$^{9}$~V, $B_{14}$ is the magnetic field in units of 10$^{14}$~G, 
$R_6$ is the NS radius in units of 10$^{6}$~cm, $\Psi$ is the magnitude of the twist in units of radians, and $u_{*}$ is the angle subtended by the arc of 
circumference of the hot spot.
The rate of decay is even faster than an exponential, since the time constant decreases as the twist angle $\Psi$ and the size of the hot spot $u_{*}$ decrease.
While this model is a good fit to the data at the beginning of \xte's outburst, this is not what has been observed in the spectra and pulse profiles from about 2007 onwards.
From 2009 March 5 through 2014 March 4 \xte\ has been in a steady state, yet the hot spot on its surface remains.

\cite{rea12} suggested that radio magnetars share the property that $L_{x} / \dot{E} <  1$, but that not all radio magnetars will necessarily satisfy this condition.
The observations of \xte \ presented here do not support this proposition.
Between 2007 and 2012 \xte \ reached a spin-down power of $\dot{E} = (5.6-6.6) \times 10^{32}$ erg s$^{-1}$ \citep{cam15}, 
while the luminosity of the cold component of \xte \ alone is $\sim 4 \times 10^{34}$ \textit{d}$^{2}_{3.5}$ erg s$^{-1}$.

\cite{sza15} explain magnetar radio emission with the partially screened gap model that has been developed to explain the radio emission of normal rotation powered pulsars.
In this scenario \xte's radio emission is also rotation powered, and radio emission is only possible if the polar cap is below the critical temperature for ion emission, and the polar cap luminosity is much less than
the spin-down power.
However, the polar cap luminosity of \xte \ increased by more than two orders of magnitude during its outburst, while the spin-down luminosity increased by a factor of 8 \citep{cam15}.
This data places a difficult constraint on how this model could explain the turn-on and turn-off of \xte's radio emission. 

\section{Conclusions}

\xte \ was the first discovered transient magnetar and therefore provides the longest record of transient magnetar behavior.
A hot spot on the NS surface is evident even in the most recent observations.
The luminosity of this hot spot exceeds \xte's spin-down power, and is therefore an indicator of continued magnetar activity.
There is currently no detailed theoretical model that explains this persistent magnetar activity.

With the benefit of over ten years of \xmm \ observations, we can say with some confidence that we have detected emission from the whole surface of \xte.
Unfortunately, large systematic uncertainties plague a measurement of the NS radius.
We nevertheless have a good measurement of the surface temperature of a magnetar.
 
The radio emission from \xte \ during its outburst is similar to the radio emission of a subgroup of the known transient magnetars.
There is no detailed theoretical model that explains the turn-on and turn-off of \xte's radio emission, but the X-ray data presented in this paper may provide an important clue.
The flux from the hot-spot on the NS surface reached its lowest level just as the radio emission disappeared.
This suggests that this magnetar's radio emission is powered by magnetic field decay.

\acknowledgements

This investigation is based on observations obtained
with \xmm, an ESA science mission with instruments and contributions
directly funded by ESA Member States and NASA.
This research made use of data obtained from the
High Energy Astrophysics Science Archive Research Center (HEASARC),
provided by NASA's Goddard Space Flight Center. We acknowledge
support from NASA ADAP grant NNX15AE63G.
We thank Eric Gotthelf for valuable discussions and the anonymous referee 
for several helpful comments.


\begin{thebibliography}{}

\bibitem[Albano et al.(2010)]{alb10} Albano, A., Turolla, R., 
Israel, G.~L., et al.\ 2010, \apj, 722, 788 

\bibitem[An et al.(2012)]{an12} An, H., Kaspi, V.~M., 
Tomsick, J.~A., et al.\ 2012, \apj, 757, 68 

\bibitem[An et al.(2013)]{an13} An, H., Kaspi, V.~M., 
Archibald, R., \& Cumming, A.\ 2013, \apj, 763, 82

\bibitem[Anderson et al.(2012)]{and12} Anderson, G.~E., 
Gaensler, B.~M., Slane, P.~O., et al.\ 2012, \apj, 751, 53 

\bibitem[Arnaud(1996)]{arn96} Arnaud, K.~A.\ 1996, in ASP Conf. Ser. 101,
Astronomical Data Analysis Software and Systems V,
ed. G. Jacoby \& J. Barnes (San Francisco: ASP), 17

\bibitem[Beloborodov(2009)]{bel09} Beloborodov, A.~M.\ 2009, 
\apj, 703, 1044 

\bibitem[Bernardini et al.(2009)]{ber09}
Bernardini, F., Israel, G.~L., Dall'Osso, S., et al.\ 2009, \aap, 498, 195 

\bibitem[Bernardini et al.(2011)]{ber11} Bernardini, F., Perna, R., Gotthelf, E. V., et al.\ 2011, \mnras, 418, 638

\bibitem[Camero et al.(2014)]{cam14} Camero, A., Papitto, A., 
Rea, N., et al.\ 2014, \mnras, 438, 3291 

\bibitem[Camilo et al.(2006)]{cam06} Camilo, F., Ransom, 
S.~M., Halpern, J.~P., et al.\ 2006, \nat, 442, 892 

\bibitem[Camilo et al.(2015)]{cam15} Camilo, F., Ransom,
S.~M., Halpern, J.~P., et al.\ 2015, \apj, submitted

\bibitem[Coti Zelati et al.(2015)]{cot15} Coti Zelati, F., 
Rea, N., Papitto, A., et al.\ 2015, \mnras, 449, 2685 

\bibitem[Demorest et al.(2010)]{dem10} Demorest, P.~B., 
Pennucci, T., Ransom, S.~M., Roberts, M.~S.~E., 
\& Hessels, J.~W.~T.\ 2010, \nat, 467, 1081 

\bibitem[Dib et al.(2012)]{dib12} Dib, R., Kaspi, V.~M., 
Scholz, P., \& Gavriil, F.~P.\ 2012, \apj, 748, 3 

\bibitem[Esposito et al.(2009)]{esp09} Esposito, P., Tiengo, 
A., Mereghetti, S., et al.\ 2009, \apjl, 690, L105 

\bibitem[Esposito et al.(2011)]{esp11} Esposito, P., Israel, 
G.~L., Turolla, R., et al.\ 2011, \mnras, 416, 205 

\bibitem[Esposito et al.(2013)]{esp13} Esposito, P., Tiengo, 
A., Rea, N., et al.\ 2013, \mnras, 429, 3123 

\bibitem[Gotthelf \& Halpern(2007)]{got07} 
Gotthelf, E.~V., \& Halpern, J.~P.\ 2007, \apss, 308, 79 

\bibitem[G{\"u}ver et al.(2007)]{guv07} G{\"u}ver, T., 
{\"O}zel, F., G{\"o}{\v g}{\"u}{\c s}, E., 
\& Kouveliotou, C.\ 2007, \apjl, 667, L73 

\bibitem[Halpern et al.(2005)]{hal05} 
Halpern, J.~P., Gotthelf, E.~V., Becker, R.~H., Helfand, D.~J., 
\& White, R.~L.\ 2005, \apjl, 632, L29 

\bibitem[Ibrahim et al.(2004)]{ibr04} Ibrahim, A.~I., 
Markwardt, C.~B., Swank, J.~H., et al.\ 2004, \apjl, 609, L21

\bibitem[Kargaltsev et al.(2012)]{kar12} Kargaltsev, O., 
Kouveliotou, C., Pavlov, G.~G., et al.\ 2012, \apj, 748, 26

\bibitem[Kaspi et al.(2014)]{kas14} Kaspi, V.~M., Archibald, 
R.~F., Bhalerao, V., et al.\ 2014, \apj, 786, 84

\bibitem[Kuiper et al.(2012)]{kui12} Kuiper, L., Hermsen, W., 
den Hartog, P.~R., \& Urama, J.~O.\ 2012, \apj, 748, 133 

\bibitem[Levin et al.(2010)]{lev10} Levin, L., Bailes, M., 
Bates, S., et al.\ 2010, \apjl, 721, L33 

\bibitem[Mereghetti et al.(2015)]{mer15} Mereghetti, S., 
Pons, J.~A., \& Melatos, A.\ 2015, \ssr, 26

\bibitem[Minter et al.(2008)]{min08} 
Minter, A.~H., Camilo, F., Ransom, S.~M., Halpern, J.~P., \& Zimmerman, N.\ 2008, \apj, 676, 1189 

\bibitem[Morrison \& McCammon(1983)]{mor83} Morrison, R., \& McCammon, D.\ 1983, \apj, 270, 119

\bibitem[Olausen \& Kaspi(2014)]{ola14} 
Olausen, S.~A., \& Kaspi, V.~M.\ 2014, \apjs, 212, 6 

\bibitem[Rea et al.(2013)]{rea13} Rea, N., Israel, G.~L., 
Pons, J.~A., et al.\ 2013, \apj, 770, 65 

\bibitem[Rea et al.(2012)]{rea12} Rea, N., Pons, J.~A., 
Torres, D.~F., \& Turolla, R.\ 2012, \apjl, 748, L12 

\bibitem[Rea et al.(2014)]{rea14} Rea, N., Vigan{\`o}, D., 
Israel, G.~L., Pons, J.~A., \& Torres, D.~F.\ 2014, \apjl, 781, L17 

\bibitem[Rybicki \& Lightman(1979)]{ryb79} 
Rybicki, G.~B., \& Lightman, A.~P.\ 1979, Radiative Processes in Astrophysics (New York: Wiley-Interscience)  

\bibitem[Scholz et al.(2014)]{sch14} Scholz, P., Kaspi, 
V.~M., \& Cumming, A.\ 2014, \apj, 786, 62 

\bibitem[Szary et al.(2015)]{sza15} 
Szary, A., Melikidze, G.~I., \& Gil, J.\ 2015, \apj, 800, 76 

\bibitem[Tam et al.(2006)]{tam06} Tam, C.~R., Kaspi, V.~M., 
Gaensler, B.~M., \& Gotthelf, E.~V.\ 2006, \apj, 652, 548 

\bibitem[Torii et al.(1998)]{tor98} Torii, K., Kinugasa, K., 
Katayama, K., Tsunemi, H., \& Yamauchi, S.\ 1998, \apj, 503, 843 

\bibitem[Vigan\`o et al.(2013)]{vig13}
Vigan\`o, D., Rea, N., \&  Pons, J.~A. 2013, \mnras, 434, 123

\bibitem[Zhou et al.(2014)]{zho14} Zhou, P., Chen, Y., Li, 
X.-D., et al.\ 2014, \apjl, 781, L16 

\end{thebibliography}
\end{document}